\documentclass[12pt]{article}
\usepackage{graphicx}
\usepackage{hfstyle}
\begin{document}

\title{Exact results for deterministic cellular automata traffic models}
\author{Henryk Fuk\'s
      \oneaddress{
         The Fields Institute for Research\\
         in Mathematical Sciences\\
         Toronto, Ontario M5T 3J1, Canada\\
         and\\
         Department of Mathematics and
         Statistics, \\University of Guelph, \\
         Guelph, Ontario N1G 2W1, Canada\\
         {~}\\
         \email{hfuks@fields.utoronto.ca}
       }
   }

\Abstract{
We present a rigorous derivation of the flow at arbitrary time in a
deterministic cellular automaton model of traffic flow. The derivation
employs regularities in preimages of blocks of zeros, reducing the
problem of preimage enumeration to a well known lattice path counting
problem. Assuming infinite lattice size and random initial
configuration, the flow can be expressed in terms of generalized
hypergeometric function. We show that the steady state limit agrees
with previously published results. }

\maketitle
\section{Introduction}
Since the introduction  of the Nagel-Schreckenberg (N-S) model in 1992
\cite{Nagel92}, cellular automata became a well established method of
traffic flow modeling. Comparatively low computational cost of
cellular automata models made it possible to conduct large-scale
real-time simulations of urban traffic in the city of Duisburg
\cite{Esser97} and Dallas/Forth Worth \cite{Simon98}. Several
simplified models have been proposed, including models based on
deterministic cellular automata. For example,  Nagel and Herrmann
\cite{Nagel93} considered deterministic version of the N-S model,
while Fukui and Ishibashi \cite{Fukui96c} introduced another model (to
be referred to as F-I model), which can be understood as a
generalization of cellular automaton rule 184. Rule 184, one of the
elementary CA rules investigated by Wolfram \cite{Wolfram94}, had been
later studied in detail as a simple model of surface growth
\cite{Krug88}, as well as in the context of density classification
problem \cite{paper4}. It is one of the only two (symmetric)
non-trivial elementary rules conserving the number of active sites
\cite{paper8}, and, therefore, can be interpreted as a rule governing
dynamics of particles (cars). Particles (cars) move to the left if
their right neighbor site is empty, and do not move if the right
neighbor site is occupied, all of them moving simultaneously at each
discrete time step. Using terminology of lattice stochastic processes,
rule 184 can be viewed as a discrete-time version of totally
asymmetric simple exclusion process. Further generalization of the F-I
model has been proposed in \cite{paper5}.

In all traffic models, the main quantity of interest is the average
velocity of cars, or the average flow, defined as a product of the
average velocity and the density of cars. The graph of the flow as a
function of density is called a fundamental diagram, and is typically
studied in the steady state ($t \rightarrow \infty$). For the F-I
model, steady-state fundamental diagram can be obtained using
mean-field argument \cite{Fukui96c}, as well as by statistical
mechanical approach \cite{Wang98b} or by studying the time evolution
of inter-car spacing \cite{Wang98a}. In general, little is known about
non-equilibrium properties of the flow. In \cite{paper4}, we
investigated dynamics of rule 184 and derived expression for the flow
at arbitrary time, assuming that the initial configuration (at $t=0$)
was random, using the concept of defects and analyzing the dynamics of
their collisions. In what follows, we shall generalize results of
\cite{paper4} for the deterministic F-I traffic flow model and derive
the expression for the flow at arbitrary time. The derivation employs
regularities of preimages of blocks of zeros, reducing the problem of
preimage enumeration to a well known combinatorial problem of lattice
path counting. Assuming infinite lattice size and random initial
configuration, the flow can then be expressed in terms of generalized
hypergeometric function. We will, unlike in \cite{paper4}, explore
regularities  of preimages using purely algebraic methods,  i.e.,
without resorting to properties of spatiotemporal diagrams and
dynamics of defects.

\section{Deterministic traffic rules}
Deterministic version of the F-I traffic model is defined on
one-dimensional lattice of $L$ sites with periodic boundary
conditions. Each site is either occupied by a vehicle, or empty. The
velocity of each vehicle is an integer between 0 and $m$. If $x(i,t)$
denotes the position of the $i$th car at time $t$, the position of the
next car ahead at time $t$ is $x(i+1,t)$. With this notation, the
system evolves according to a synchronous rule given by
\begin{equation}
x(i,t+1)=x(i,t)+v(i,t),
\label{eq1}
\end{equation}
where
\begin{equation}
 v(i,t)=\min\big(x(i+1,t)-x(i,t)-1,m\big)
 \label{eq2}
\end{equation}
is the velocity of car $i$ at time $t$. Since $g=x(i+1,t)-x(i,t)-1$ is
the gap (number of empty sites) between cars $i$ and $i+1$ at time
$t$, one could say that each time step, each car advances by $g$ sites
to the right if $g \leq m$, and by $m$ sites if $g>m$.  When $m=1$,
this model is equivalent to elementary cellular automaton rule 184,
for which a number of exact results is known~\cite{Krug88,paper4}.

The main quantities of interest in this paper will be the average
velocity of cars at time $t$ defined as
\begin{equation}
  \overline{v}(t)=\frac{1}{N} \sum_{i=1}^N v(i,t),
\end{equation}
and the average flow $\phi(t)=\rho \overline{v}(t)$, where $\rho=N/L$
is the density of cars. In what follows, we will assume that at $t=0$
the cars are randomly distributed on the lattice. When $N \rightarrow
\infty$, this corresponds to a situation when sites are occupied
by a car with probability $\rho$, or are empty with probability
$1-\rho$.

In general, if $N_k(t)$ is the number of cars with velocity $k$, we
have
\begin{equation} \label{vn}
  \overline{v}(t)=\frac{1}{N} \sum_{k=1}^m k N_k(t).
\end{equation}
When $k<m$, $N_k(t)$ is just the number of blocks of type $10^k1$,
where $0^k$ denotes $k$ zeros. This means that a probability of an
occurrence of the block $10^k1$ at time $t$ can be written as
$P_t(10^k1)=N_k/L$. Similarly, for $k=m$, $P_t(10^m)=N_m(t)/L$. As a
consequence, equation
 (\ref{vn}) becomes
 \begin{equation} \label{vp}
  \overline{v}(t) =  \sum_{k=1}^{m-1} \frac{k P_t(10^k1)}{\rho} +
  \frac{m P_t(10^m)}{\rho}
 \end{equation}
We will now demonstrate that in the deterministic F-I model with
maximum speed $m$ the average flow depends only on one block
probability. More precisely, we shall prove the following: \\ {\bf
Proposition 1. }{\it In the deterministic F-I model with the maximum
 speed $m$, the average flow $\phi_m(t)$ is given by
 \begin{equation} \label{prop1}
  \phi_m(t)=1-\rho - P_t(0^{m+1}).
 \end{equation}
}

To prove this proposition by induction, we first note that for $m=1$
equation (\ref{vp}) gives $\phi_1(t)=P_t(10)$. Using consistency
condition for block probabilities $P_t(10)+P_t(00)=P_t(0)=1-\rho$, we
obtain $\phi_1(t)=1-\rho - P(00)$, which verifies (\ref{prop1}) in the
$m=1$ case. Now assume that (\ref{prop1}) is true for some $m=n-1$
(where $n>1$), and compute $\phi_{n}(t)$:
\begin{eqnarray*}
\phi_{n}(t)&=&nP_t(10^n) + \sum_{j=1}^{n-1} j P(10^j1)=\\
&=& nP_t(10^n) +(n-1) P_t(10^{n-1}1)+ \sum_{j=1}^{n-2} j P(10^j1)\\
&=&(n-1)[P_t(10^{n-1}1)+P_t(10^n)] +P_t(10^n) + \sum_{j=1}^{n-2} j
P(10^j1)
\end{eqnarray*}
Using consistency condition $P_t(10^{n-1}1)+P_t(10^n)=P_t(10^{n-1})$
we obtain
\begin{eqnarray*}
\phi_{n}(t)=P_t(10^n) + (n-1)P_t(10^{n-1})  +
\sum_{j=1}^{n-2} j P(10^j1)=P_t(10^n)+\phi_{m-1}(t)
\end{eqnarray*}
Taking into account that $P_t(10^n)=P_t(0^n)-P_t(0^{n+1})$ (which,
again, is just a consistency condition for block probabilities), and
using (\ref{prop1}) to express $\phi_{m-1}(t)$, we finally obtain
\begin{equation}
  \phi_m(t)=1-\rho - P_t(0^{m+1}).
 \end{equation}
 This means that validity of (\ref{prop1}) for $m=n$ follows from
 its validity for $m=n-1$, concluding our proof by induction.

\section{Enumeration of preimages of $0^{m+1}$}
Proposition~1 reduces the problem of computing $\phi_m(t)$ to the
problem of finding the probability of a block of $m+1$ zeros.  In
order to find this probability, we will now use the fact that the
deterministic F-I model is equivalent to a cellular automaton defined
as follows. Let $s(i,t)$ denotes the state of a lattice site $i$ at
time $t$ (note that $i$ now labels consecutive lattice sites,
\emph{not} consecutive cars), where $s(i,t)=1$ for a site occupied by
a car and $s(i,t)=0$ otherwise. We can immediately realize that if a
site $i$ is empty at time $t$, then at time $t+1$ it can become
occupied by a car arriving from the left, but not from a site further
than $i-m$. Similarly, if a site $i$ is occupied, it will become empty
at the next time step only and only if site $i+1$ is empty. Thus, in
general, $s(i,t+1)$ depends on $s(i-m,t), s(i-m+1,t), \ldots,
s(i+1,t)$, i.e., on the state of $m$ sites to the left, one site to
the right, and itself, but not on any other site, what can be
expressed as
\begin{equation}
 s(i,t+1)=f_m\Big(s(i-m,t), s(i-m+1,t), \ldots, s(i+1,t) \Big),
\end{equation}
where $f_m$ is called a local function of the cellular automaton. For
the F-I CA, one can write explicit formula\footnote{Since formula
(\ref{explicit}) will not be used in subsequent calculations, we give
it withot proof  (which is elementary).}  for $f_m$, such as
\begin{eqnarray} \label{explicit}
 f_m\Big(s(i-m,t), s(i-m+1,t), \ldots, s(i+1,t) \Big)= s(i,t)-
\min\{s(i,t),1-s(i+1,t)\} \nonumber \\
+\min\Big\{\max\{s(i-m,t),s(i-m+1,t),\ldots,s(i-1,t)\} ,
1-s(i,t)\Big\},
\end{eqnarray}
which, using terminology of cellular automata theory, represents a
rule with left radius $m$ and right radius $1$. In general, after $t$
iteration of this cellular automaton rule, state of a site $s(i,t)$
depends on $s(i-mt,0), s(i-mt+1,0),
\ldots , s(i+t,0)$, but not on any other sites in the initial
configuration. Similarly, a block of $k$ sites $s(i,t) s(i+1,t) \ldots
s(i+k)$ depends only on a block $s(i-mt,0), s(i-mt+1,0), \ldots ,
s(i+k+t,0)$, as schematically shown in Figure 1.
\begin{figure}
\begin{center}
\includegraphics[scale=0.8]{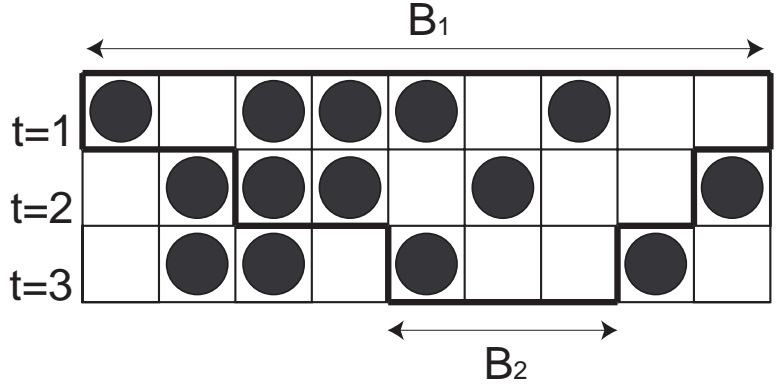}
\end{center}
\caption{Fragment of a spatiotemporal diagram for the F-I rule with $m=2$.
States of nine sites during three consecutive time steps are are
shown, black circles representing occupied sites. Block
$B_1=101110100$ is a $2$-step preimage of the block $B_2=100$.
Outlined sites constitute ``light cone'' of the block $B_2$, meaning
that the state of sites belonging to $B_2$ can depend only on sites
inside the outlined region, but not on sites outside this region.}
\end{figure}
 We will say that $s(i-mt,0), s(i-mt+1,0), \ldots ,
s(i+k+t,0)$ is an $t$-step preimage of the block $s(i,t) s(i+1,t)
\ldots s(i+k)$. Preimages in the F-I cellular automaton have the following
property:

\noindent {\bf Proposition 2.} {\em Block $a_1a_2a_3 \ldots a_p$ is an n-step
preimage of a block $0^{m+1}$ if and only if  $p=(n+1)(m+1)$ and, for
every $k$ ($1 \leq k \leq p$)
\begin{equation} \label{condxi}
\sum_{i=1}^k \xi(a_i)>0,
\end{equation}
where $\xi(1)=-m$ and $\xi(0)=1$.}

Before we present a proof of this proposition, note that it can be
interpreted as follows. Let us assume that we have a block of zeros
and ones of length $p$, where $p=(n+1)(m+1)$, and we want to check if
this block is an $n$-step preimage of a block $0^{m+1}$. We start with
a ``capital'' equal to zero. Now we move from the leftmost site to the
right, and every time we encounter $0$, we increase our capital by
$m$. Every time we encounter $1$, our capital decreases by $1$. If we
can move from $a_1$ to $a_p$ and our capital stays always larger than
zero, the string $a_1a_2a_3
\ldots a_p$ is a preimage of $0^{m+1}$.
Condition (\ref{condxi}) can be also written as
\begin{equation} \label{conda}
\sum_{i=1}^k a_i < \frac{k}{m+1},
\end{equation}
because $\xi(x)=1-(m+1)x$ for $x \in \{0,1\}$.

For the purpose of the proof, strings  $a_1a_2 \ldots a_p$ of length
$p$ satisfying (\ref{conda}) for a given $m$ and for every $k\leq N$
will be called {\em $m$-admissible strings}.

\noindent {\bf Lemma.} {\em Let $s(1,t)s(2,t) \ldots s(p,t)$ be an
$m$-admissible string. If
\begin{equation}
 s(i,t+1)=f_m\Big(s(i-m,t), s(i-m+1,t), \ldots, s(i+1,t) \Big),
\end{equation}
and if $f_m$ is a local function of the deterministic F-I model with
maximum speed $m$, then $s(m+1,t+1) s(m+2,t+1) \ldots s(p-1,t+1)$ is
also an $m$-admissible string.}

To prove the lemma, it is helpful to employ the fact that the F-I rule
conserves the number of cars. Let $0<k<p$ and let us consider strings
$S_1=s(1,t)s(2,t) \ldots s(k,t)$ and $S_2=s(m+1,t+1)s(2,t+1) \ldots
s(k,t+1)$. If the string $s(1,t)s(2,t) \ldots s(k,t)$ is
$m$-admissible, then its first $m+1$ sites must be zeros. This means
that in one time step, no car can enter  string $s(1,t)s(2,t) \ldots
s(k,t)$ from the left. On the other hand, in a single time step, only
one car (or none) can leave the string on the right hand side, i.e.,
\begin{equation}
\sum_{i=1}^k s(i,t) =  \epsilon + \sum_{i=m+1}^k s(i,t+1)
\end{equation}
where $\epsilon \in \{0,1\}$. Three cases can be distinguished:

{\bf (i)} All sites $s(k-m+1,t) s(k-m+2,t) \ldots s(k,t)$ are empty
(equal to~$0$). Then no car leaves $S_1$, which means that
$\epsilon=0$, and
\begin{equation}
\sum_{i=1}^k s(i,t) = \sum_{i=m+1}^k s(i,t+1) = \sum_{i=m+1}^{k-m} s(i,t+1)
< \frac{k-m}{m+1}.
\end{equation}
The last inequality is a direct consequence of $m$-admissibility of
$S_1$. Since the length of the string $S_2$ is equal to $k-m$, the
above relation (which holds for arbitrary $k$) proves that $S_2$ is
also $m$-admissible in the case considered.

{\bf (ii)} Among sites $s(k-m+1,t) s(k-m+2,t) \ldots s(k,t)$ there is
at
 least one which is occupied (equal to 1), and $s(k+1,t)=1$. In this case,
 since the last site in $S_1$ is ``blocked'' by the car at $s(k+1,t)$,
 again no car can leave string $S_1$ in one time step. Therefore,
\begin{equation} \label{iia}
 \sum_{i=1}^k s(i,t) = \sum_{i=m+1}^k s(i,t+1).
\end{equation}
$m$-admissibility of $S_2$ implies
\begin{equation} \label{iib}
\frac{k+1}{m+1} > \sum_{i=1}^{k+1}s(i,t)=\sum_{i=1}^{k}s(i,t) +1.
\end{equation}
Combining (\ref{iia}) with (\ref{iib}) we obtain
\begin{equation}
 \sum_{i=m+1}^{k-m} s(i,t+1) < \frac{k-m}{m+1},
\end{equation}
which again shows that $S_2$ is $m$-admissible.

{\bf (iii)} Among sites $s(k-m+1,t) s(k-m+2,t) \ldots s(k,t)$ there is
at
 least one which is occupied (equal to 1), and $s(k+1,t)=0$. In this case,
 one car will leave right end of the string $S_1$, therefore
\begin{equation}
 \sum_{i=1}^k s(i,t) = \sum_{i=m+1}^k s(i,t+1) -1.
\end{equation}
As before, from $m$-admissibility of $S_1$ we have
\begin{equation}
\sum_{i=1}^{k+1}s(i,t)=\sum_{i=1}^{k}s(i,t)<\frac{k+1}{m+1},
\end{equation}
hence
\begin{equation}
\sum_{i=m+1}^k s(i,t+1)=\sum_{i=m+1}^k s(i,t+1) -1<\frac{k+1}{m+1} -1
= \frac{k-m}{m+1},
\end{equation}
which demonstrates that case (iii) also leads to $m$-admissibility of
$S_2$, concluding the proof of our lemma.

Let us  now assume that the block $B_1=s(1,t) s(2,t)
\ldots s(p,t)$ is $m$ admissible ($n$ being some
fixed integer and $p=(m+1)(n+1)$). Applying the lemma to this block we
conclude that $B_2=s(m+1,t+1) s(2,t+1)\ldots s(p-1,t+1)$ is
$m$-admissible as well. Applying the lemma to $B_2$ we obtain
$m$-admissible block  $B_3=s(2m+1,t+2) s(2,t+2)\ldots s(p-2,t+2)$.
After $n$ applications of the lemma we end up with the conclusion that
the string $B_{n+1}=s(nm+1,n+1)s(nm+2,n+1) \ldots s(p-n)$ is
$m$-admissible. Since the length of $B_{n+1}$ is $p-n -
nm=(n+1)(m+1)-n(m+1)=m+1$, it must, to be $m$-admissible, be composed
of all zeros, i.e., $B_{m+1}=0^{m+1}$. This means that
$m$-admissibility of $B_1$ is a sufficient condition for $B_1$ to  be
an $n$-step preimage of $0^{m+1}$. Reversing steps in the above
reasoning, one can show that is is also a necessary condition.

\section{Fundamental diagram}
We shall now use proposition 2 to calculate $P_t(0^{m+1})$. First of
all, we note that $P_t(0^{m+1})$ is equal to the probability of
occurrence of $t$-step preimage of $0^{m+1}$ in the initial (random)
configuration, that is
\begin{equation}
P_t(0^{m+1})=\sum P_0(a),
\end{equation}
where the sum goes over all $t$-step preimages of $0^{m+1}$. Consider
now a string which contains $n_0$ zeros and $n_1$ ones. The number of
such strings can be immediately obtained if we realize that  it is
equal to the number of lattice paths from the origin to $(n_0,n_1)$
which do not touch nor cross the line $x=my$, as shown in Figure 2.
\begin{figure}
\begin{center}
\includegraphics[scale=0.6]{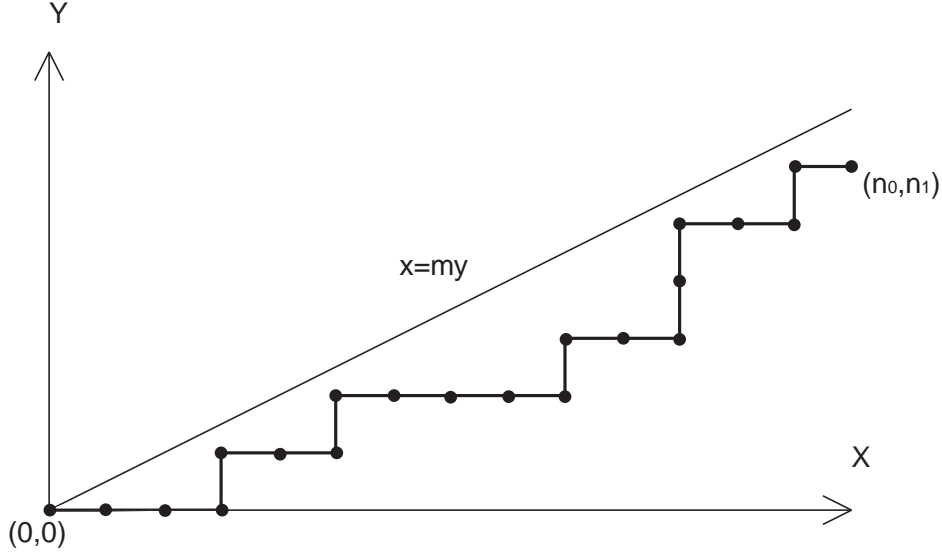}
\end{center}
\caption{$m$-admissible  block  with $n_0$ zeros and $n_1$ ones
is equivalent to a lattice path from the origin to $(n_0,n_1)$ which
does not touch nor cross the line $x=my$.  0 corresponds to a
horizontal segment, while 1 to a vertical segment.}
\end{figure}
This is  a well known combinatorial problem \cite{Mohanty79}, and the
number of aforementioned paths equals
\begin{equation}
\frac{n_0 - m n_1}{n_0+n_1} {{n_0+n_1} \choose n_1}
\end{equation}
Probability of occurrence of such a block in a random configurations
is, therefore,
\begin{equation}
\frac{n_0 - m n_1}{n_0+n_1} {{n_0+n_1} \choose n_1} \rho^{n_1}
(1-\rho)^{n_0},
\end{equation}
where $\rho=P(1)$. In a $t$-step preimage of $0^{m+1}$ the minimum
number of zeros is $ 1+ m(t+1)$ zeros, while the maximum is
$(m+1)(t+1)$ (corresponding to all zeros). Therefore, summing over all
possible number of zeros $i$, we obtain
\begin{eqnarray*}
 P_t(0^{m+1})&=&\sum_{i=1+m(t+1)}^{(m+1)(t+1)}
 \frac{i-m[(m+1)(t+1)-i]}{(m+1)(t+1)}
 {{(m+1)(t+1)} \choose {(m+1)(t+1)-i}} \\
 && \times \rho^{(m+1)(t+1)-i} (1-\rho)^i
\end{eqnarray*}
Changing summation index $j=i-m(t+1)$ we obtain
\begin{equation} \label{maineq}
P_t(0^{m+1}) = \sum_{j=1}^{t+1} \frac{j}{t+1} {{(m+1)(t+1)} \choose
 {t+1-j}} \rho^{t+1-j} (1-\rho)^{m(t+1)+j}.
\end{equation}
\begin{figure}
\begin{center}
\includegraphics[scale=0.9]{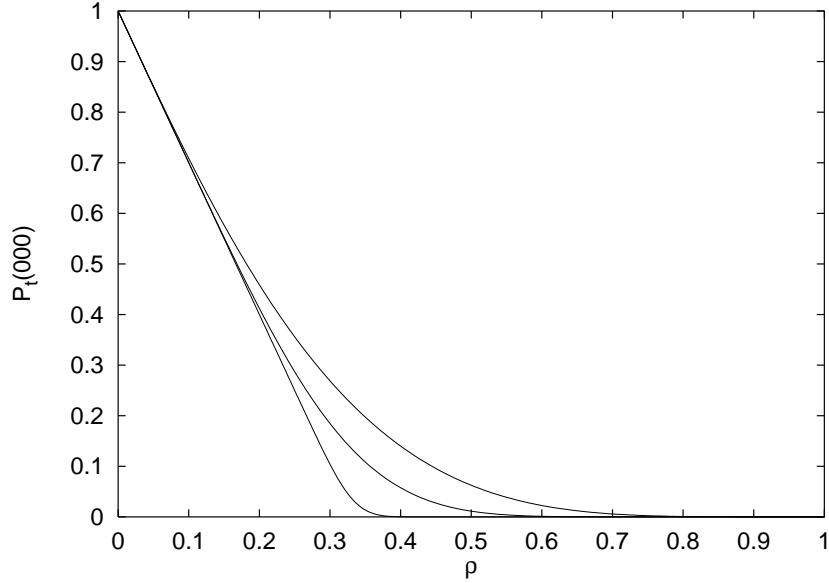}
\end{center}
\caption{Graph of the probability $P_t(0^{m+1})$ as a function of $\rho$
for $m=2$ and $t=1$ (upper line), $t=5$ (middle line), and $t=100$
(lower line).}
\end{figure}
Figure 3. shows a graph of $P_t(0^{m+1})$ as a function of $\rho$ for
$m=2$ and several values of $t$. We can observe that as $t$ increases,
the graph becomes ``sharper'' at $\rho=1/3$, eventually developing
singularity (discontinuity in the first derivative) at $\rho=1/3$.
More precisely, one can show (see appendix) that
\begin{equation} \label{plimit}
 \lim_{t \rightarrow \infty} P_t(0^{m+1})=
 \left\{ \begin{array}{ll}
 1-(m+1)\rho  & \mbox{if $p<1/(m+1)$}, \\
 0    & \mbox{otherwise}.
\end{array}
\right.
\end{equation}
$P_\infty(0^{m+1})$, therefore, can be viewed as the order parameter
in a phase transition with critical point at $\rho=1/(m+1)$. Using
Proposition~1 we can now find the average flow in the steady state
\begin{equation}
\phi_m(\infty)= \left\{ \begin{array}{ll}
 m \rho  & \mbox{if $p<1/(m+1)$}, \\
 1-\rho    & \mbox{otherwise},
\end{array}
\right.
\end{equation}
which agrees  with mean-field type calculations reported in
\cite{Fukui96c} as well as with results of \cite{Wang98b,Wang98a}.

To verify validity of the result for $t<\infty$, we performed computer
simulations using a lattice of $10^5$ sites with periodic boundary
conditions. The average flow has been recorded after each iteration up
to $t=100$ for three values of $\rho$: at the critical point
$\rho=1/3$ as well as below and above the critical point. The
resulting plots of the flow as a function of time are presented in
Figure 4. Again, the agreement with theoretical curves
\begin{equation} \label{normform}
\phi_m(t) =1-\rho- \sum_{j=1}^{t+1} \frac{j}{t+1} {{(m+1)(t+1)} \choose
 {t+1-j}} \rho^{t+1-j} (1-\rho)^{m(t+1)+j}
\end{equation}
is very good.

Without going into details, we note that the formula (\ref{normform})
can be also expressed in terms of generalized hypergeometric function
${_2{\rm F}_1}$:
\begin{eqnarray} \label{hypform}
\phi_m(t) =1-\rho-
 \frac{( 1 - \rho )^{1 + m + m t} \rho^t (1 + m + t + m t)!}
 {( 1 + m + m t) (1 + t)!(m + m t)!}
 \; {_2{\rm F}_1} \!\!\left[ \begin{array}{c}
 2 \, , \, -t \\ 2 + m + m t
 \end{array} ; 1 - \frac{1}{\rho} \right],
\end{eqnarray}
Since fast numerical algorithms for computing ${_2{\rm F}_1}$ exist,
this form might be useful for the purpose of numerical evaluation of
$\phi_m(t)$.
\begin{figure}
\begin{center}
\includegraphics[scale=0.9]{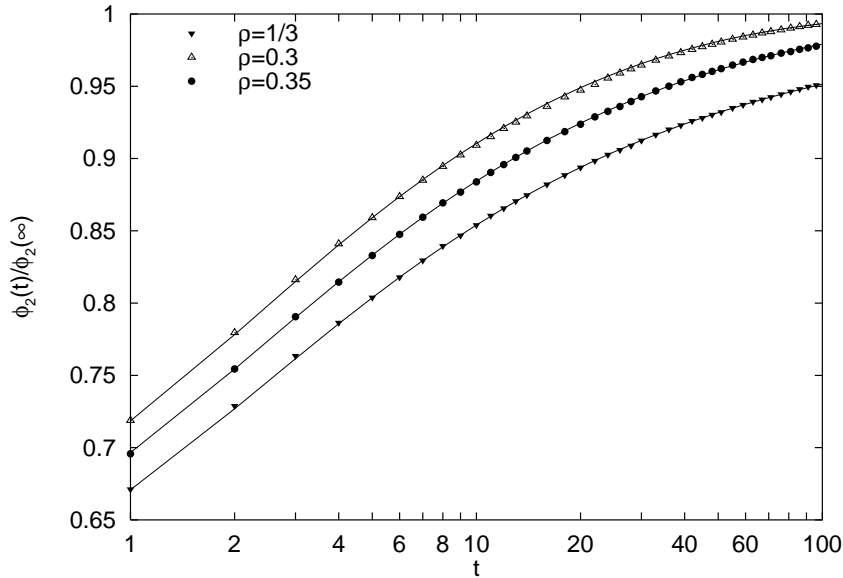}
\end{center}
\caption{Plots of $\phi_2(t)/\phi_2(\infty)$ as a function of time for
$\rho=0.3$, $\rho=1/3$, and  $\rho=0.35$ obtained from computer
simulation on a lattice of $10^5$ sites. Continuous line corresponds
to the theoretical result obtained using eq.~(\ref{hypform}).}
\end{figure}

\section{Conclusion}
We presented derivation of the flow at arbitrary time in the
deterministic F-I cellular automaton model of traffic flow. First, we
showed that the flow can be expressed by the probability of occurrence
of the block of $m+1$ zeros $P(0^{m+1})$. By employing regularities in
preimages of blocks of zeros, we reduced the problem of preimage
enumeration to the lattice path counting problem. Finally, we used the
number of preimages to find $P(0^{m+1})$, which determines the flow.

We also found that the flow in the steady state, obtained by taking $t
\rightarrow \infty$ limit, agrees with previously reported mean-field
type calculations, meaning that in the case of the F-I model
mean-field approximation gives exact results. This seems to be true
not only for the F-I model, but also for many other CA rules
conserving the number of active sites (``conservative'' CA). For
example, in \cite{paper8} we reported that the third order local
structure approximation, which is a generalization of simple
mean-field theory incorporating short-range correlations, yields the
fundamental diagram for rule 60200 (one of the $4$-input
``conservative'' CA rules) in extremely good agreement with computer
simulations. Taking this into account, we conjecture that the local
structure approximation gives exact fundamental diagram for almost all
``conservative'' rules, excluding, perhaps, those rules for which the
fundamental diagram is not sufficiently ``regular'' (meaning not
piecewise linear). This problem is currently under investigation.

\section*{Acknowledgements}
The author wishes to thank The Fields Institute for Research in
Mathematical Sciences for generous hospitality  and the Natural
Sciences and Engineering Research Council of Canada for financial
support.

\section*{Appendix}
In order to find the limit $\lim_{t \rightarrow
\infty} P_t(0^{m+1})$ we can write eq. (\ref{maineq}) in the form
\begin{equation} \label{binomform}
P_t(0^{m+1})=\sum_{j=1}^{t+1}\frac{j}{t+1} b(t+1-j,(m+1)(t+1),\rho),
\end{equation}
where
\begin{equation}
b(k,n,p)={n \choose k} p^k (1-p)^{n-k}
\end{equation}
is the distribution function of the binomial distribution. Using
de~Moivre-Laplace limit theorem, binomial distribution for large $n$
can be approximated by the normal distribution
\begin{equation} \label{demoivre}
b(k,n,p)\sim\frac{1}{\sqrt{2 \pi np(1-p)}}
\exp{\frac{-(k-np)^2}{2np(1-p)}}.
\end{equation}
To simplify notation, let us define $T=t+1$ and $M=m+1$. Now, using
(\ref{demoivre})  to approximate $b(T-j,M T,\rho)$ in
(\ref{binomform}), and approximating sum by an integral, we obtain
\begin{equation}
P_{t}(0^{m+1})=\int_{1}^{T} \frac{x}{T}
\frac{1}{\sqrt{2 \pi MT\rho(1-\rho)}}
\exp{\frac{-(T-x-M T \rho)^2}{2 M T \rho(1-\rho)}} dx.
\end{equation}
Integration yields
\begin{eqnarray*}
P_{t}(0^{m+1})= \sqrt{\frac{M \rho (1-\rho)}{2 \pi T}}
\left\{\exp\left(\frac{-(1-T+M \rho T)^2}{2 M T
\rho (1-\rho)}\right) - \exp\left({\frac{-M \rho T
}{2 (1-\rho)}}\right) \right\} +\\
\frac{1}{2}(1-M\rho)  \left\{
{\rm erf}\left(\frac{M \rho T}{\sqrt{2 M \rho (1-\rho) T}}\right) -
{\rm erf}\left(\frac{1- T + M \rho T}{\sqrt{2 M \rho (1-\rho)
T}}\right)
\right\},
\end{eqnarray*}
where ${\rm erf}(x)$ denotes the error function
\begin{equation}
{\rm erf(x)}=\frac{2}{\sqrt{\pi}} \int_{0}^{x} e^{-t^2} dt.
\end{equation}
 The first term in the
above equation (involving two exponentials) tends to~$0$ with~$T
\rightarrow
\infty$. Moreover, since $\lim_{x \rightarrow \infty}{\rm erf}(x)=1$,
 we obtain
\begin{eqnarray*}
\lim_{t \rightarrow
\infty} P_t(0^{m+1})=\frac{1}{2}(1-M\rho)
\left\{ 1-
\lim_{T \rightarrow \infty} {\rm erf}\left(\frac{1- T +
M \rho T}{\sqrt{2 M \rho (1-\rho)
T}}\right)
\right\}.
\end{eqnarray*}
Now, noting that
\begin{equation}
\lim_{T \rightarrow \infty}
{\rm erf}\left(\frac{1- T + M \rho T}{\sqrt{2 M \rho (1-\rho)
T}}\right)=
\left\{ \begin{array}{ll}
 1,  & \mbox{if $M \rho \geq 1$}, \\
 -1,    & \mbox{otherwise},
\end{array}
\right.
\end{equation}
and returning to the original notation, we recover eq. (\ref{plimit}):
\begin{equation}
 \lim_{t \rightarrow \infty} P_t(0^{m+1})=
 \left\{ \begin{array}{ll}
 1-(m+1)\rho  & \mbox{if $p<1/(m+1)$}, \\
 0    & \mbox{otherwise}.
\end{array}
\right.
\end{equation}


\begin{thebibliography}{10}

\bibitem{paper8}
Nino Boccara and Henryk Fuk{\'s}.
\newblock Cellular automaton rules conserving the number of active sites.
\newblock {\em J. Phys. A: Math. Gen.}, 31:6007--6018, 1998, adap-org/9712003.

\bibitem{Esser97}
J.~Esser and M.~Schreckenberg.
\newblock Microscopic simulation of urban traffic based on cellular automata.
\newblock {\em Int. J. Mod. Phys. C}, 8:1025--1036, 1997.

\bibitem{paper4}
Henryk Fuk{\'s}.
\newblock Solution of the density classification problem with two cellular
  automata rules.
\newblock {\em Phys. Rev. E}, 55:2081R--2084R, 1997, comp-gas/9703001.

\bibitem{paper5}
Henryk Fuk{\'s} and Nino Boccara.
\newblock Generalized deterministic traffic rules.
\newblock {\em Int. J. Mod. Phys. C}, 9:1--12, 1998, adap-org/9705003.

\bibitem{Fukui96c}
M.~Fukui and Y.~Ishibashi.
\newblock Traffic flow in {1D} cellular automaton model including cars moving
  with high speed.
\newblock {\em J. Phys. Soc. Japan}, 65:1868--1870, 1996.

\bibitem{Krug88}
J.~Krug and H.~Spohn.
\newblock Universality classes for deterministic surface growth.
\newblock {\em Phys. Rev. A}, 38:4271--4283, 1988.

\bibitem{Mohanty79}
Sri~Gopal Mohanty.
\newblock {\em Lattice Path Counting and Applications}.
\newblock Academic Press, New York, 1979.

\bibitem{Nagel92}
K.~Nagel and M.~Schreckenberg.
\newblock A cellular automaton model for freeway traffic.
\newblock {\em J. Physique I}, 2:2221--2229, 1992.

\bibitem{Nagel93}
Kai Nagel and Hans~J. Herrmann.
\newblock Deterministic models for traffic jams.
\newblock {\em Physica A}, 199:254--269, 1993.

\bibitem{Simon98}
P.~M. Simon and K.~Nagel.
\newblock Simplified cellular automaton model for city traffic.
\newblock {\em Phys. Rev. E}, 58:1286--1295, 1998, cond-mat/9801022.

\bibitem{Wang98b}
Bing-Hong Wang, Yvonne-Roamy Kwong, and Pak-Ming Hui.
\newblock Statistical mechanical approach to {F}ukui-{I}shibashi traffic flow
  models.
\newblock {\em Phys. Rev. E}, 57:2568--2573, 1998.

\bibitem{Wang98a}
Bing-Hong Wang, Lei Wang, P.~M. Hui, and Bambi Hu.
\newblock Analytical results for steady state of traffic flow models with
  stochastic delay.
\newblock {\em Phys. Rev. E}, 58:2876--2882, 1998, cond-mat/9804269.

\bibitem{Wolfram94}
S.~Wolfram.
\newblock {\em Cellular Automata and Complexity: Collected Papers}.
\newblock Addison-Wesley, Reading, Mass., 1994.

\end{thebibliography}
\end{document}